\documentclass[aps,final,notitlepage,oneside,twocolumn,nobibnotes,nofootinbib,
superscriptaddress,noshowpacs,centertags]{revtex4-1}

\usepackage[utf8]{inputenc}
\usepackage[english]{babel}
\usepackage{graphicx}
\usepackage{latexsym}
\usepackage{amssymb}
\usepackage{amsmath}
\usepackage{float}

\begin{document}

\title{Capture of the free-floating planets and primordial black holes into protostellar clouds}

\author{Yury N. Eroshenko}\thanks{e-mail: eroshenko@inr.ac.ru}
\affiliation{Institute for Nuclear Research of the Russian Academy of Sciences, Moscow 117312, Russia}

\date{\today}

\begin{abstract}
The capture of the free-floating planets and primordial black holes into a collapsing protostellar cloud is considered. Although the last stage of rapid contraction leading to the star formation lasts for a relatively short time $\sim 10^5$~years, during this time there is a strong change in the gravitational potential created by the movement of the entire cloud's mass ($\sim M_\odot$). As a result, the probability of capturing an object into a contracting cloud is comparable to the probability of capturing into an already formed planetary system. Taking into account the collapse of the cloud increases by 70\% the full probability of the planets capture at the orbits with large semi-axis $a<10^3$~au. Capture in the cloud can explain the wide inclined orbit of the supposed 9th planet in the solar system. At the same time, the probability of primordial black holes capturing from the galactic halo into a contracting cloud is extremely small.
\end{abstract}

\maketitle 




\section{Introduction}

In recent years, thousands of planets in orbits around other stars (exoplanets) have been discovered by various methods. In addition to the planets near stars, microlensing OGLE (Optical Gravitational Lensing Experiment) data indicates the existence of the free-floating planet population \cite{Mroetal17}. The presence of this population is quite natural, since planetary systems can be subjected to tidal gravitational influence from nearby stars, especially if the host star is in an association or cluster. The presence in the association is most likely at the initial stage of the star's life in the star formation region. According to modern concepts, the formation of most stars occurs in molecular clouds due to fragmentation and gravitational instability \cite{Sur01}, therefore, the star formation region may contain hundreds or thousands of stars. Subsequently, such stellar associations can dissipate \cite{Tut78}. It is also possible to release planets and comets from the protoplanetary disks \cite{TutDreDre20} or at some later stages of the planetary system evolution \cite{AndPop21}.  

There are many exoplanets at very wide orbits or with large inclinations \cite{PerKou12}. Since these planets, apparently, could not form in such orbits, the hypothesis was proposed that either they were ejected from the inner part of the planetary system, or were born at other stars, were lost, and then captured by another star. The capture mechanism considered was the interaction of a passing planet with a system of two bodies: a star and planet in orbit around the star \cite{VarSgaTsi12,GouRib17}. The capture process is the reverse process with respect to the slingshot ejection \cite{NapAdaBat21}. In the solar system, the capture of bodies is most likely due to the variable gravitational field of the Sun-Jupiter system. Like the tidal loss of planets, the capture process is most likely in the region of star formation, where there should be a large number density of free-floating planets lost by other stars \cite{PerKou12,MusRayDav16,ParLicQua17,DafParQua22}. 

In the works \cite{BatBro16,BatBro19}, indications were obtained that the orbits of trans-Neptunian objects are correlated in a certain way, which may point to the presence of 9th planet with a mass $5-10M_\oplus$ on the periphery of the solar system. The probable parameters of its orbit are: the semimajor axis $a=400-1000$~au, the eccentricity $e=0.2-0.5$ and the inclination $30-60^\circ$. A planet with similar orbital parameters could also be captured in the solar system \cite{MusRayDav16,LiAdam16}. In the work \cite{SchUnw20}, the hypothesis was proposed that the 9th planet is a captured primordial black hole (PBH).

In the above works, the capture into an already formed planetary system was considered. In this paper, we point out the existence of another (additional) capture mechanism at the stage before the formation of a planetary system and even before the formation of a star. We are considering the capture of free-floating planets in a contracting protostellar cloud. When a planet passes by a static gravitating object, the planet first gains kinetic energy, and then loses exactly the same kinetic energy. If a planet flies into a contracting protostellar cloud, then when moving to the center of the cloud, a smaller force of attraction acts on the planet than when moving further away from the center, because the density of the cloud increases over time, and the force of gravity at a fixed radial distance from the center increases with time. As a result, a planet floating into the cloud at the stage of approaching the center of the cloud acquires less kinetic energy than it loses at a subsequent fly out from the center. With a sufficiently large difference between the lost and acquired energies, the planet may end up in a finite elliptical orbit. 

In this paper, we performed statistical modelling and found the probability of capturing planets in a contracting cloud. With certain plausible assumptions about the structure of the cloud and the structure of the surrounding region of star formation, it turns out that the probability of capture into the cloud is comparable to the probability of capture into an already formed planetary system. Due to this, the overall probability of capturing the planets increases by about 70\%. We are also consider the capture of PBHs from the Galactic halo into a contracting cloud, assuming that PBHs make up a certain fraction of dark matter in the halo. The probability of PBH capture is much less in comparison with the planet capture  in a star formation region, because the velocity dispersion of the halo objects is much larger.


\section{Capture dynamics into a contracting cloud}

Consider a protostellar molecular cloud with mass $M\sim M_\odot$ at the stage of its evolution, when the cloud loses stability and begins to contract in free fall mode. Let this stage of compression begin at the time $t_i$ (further, the index ``i'' at variables denotes their values at the time $t_i$).  Typical cloud parameters at the beginning of this stage are given in \cite{Sur01}. The temperature of the gas in the cloud is $\simeq10$~K, the number density of hydrogen molecules is $n_i\simeq10^5$~cm$^{-3}$ and, accordingly, the density of the cloud is $\rho_i\simeq 2m_pn_i$, where $m_p$ is the mass of the proton (the contribution of helium is neglected), the Jeans radius is  $4\times10^3$~au, while the size of the cloud itself is $R_i=7.5\times10^3$~au. We assume that the cloud is approximately homogeneous and spherically symmetric. A strong inhomogeneity with an increase in density towards the center develops at the very last stage of compression, when the cloud already becomes opaque. 

The radius of a cloud contracting from a state of rest changes according to the equation 
\begin{equation}
\frac{d^2R}{dt^2}=-\frac{GM}{R^2},
\end{equation}
whose solution can be written in parametric form
\begin{equation}
R=R_i\cos^2\theta,
\label{rparam}
\end{equation}
\begin{equation}
t=t_i+\left(3/2\right)^{1/2}t_d\left[\theta+\frac{1}{2}\sin\left(2\theta\right)\right],
\end{equation}
where the initial dynamic time is
\begin{equation}
t_d=\frac{1}{\sqrt{4\pi G\rho_i}}\simeq6\times10^4\left(\frac{n_i}{10^5\mbox{cm$^{-3}$}}\right)\mbox{~yr}.
\end{equation}
The compression time when changing the $\theta$ parameter from $0$ to $\pi/2$ is $(3/2)^{1/2}t_d(\pi/2)\simeq1.9t_d$. The density of the contracting cloud is
\begin{equation}
\rho(t)=\rho_i\frac{R_i^3}{R^3(t)}.
\end{equation}

Consider an object that at the initial moment has a radius vector $\vec r_i$ with respect to the center of the cloud. This object at the time $t_i$ could be either inside or outside the cloud. Denote by $\gamma_i$ the angle between $\vec r_i$ and the initial velocity direction $\vec v_i$, then the specific angular momentum of the object $l=r^2\dot\phi=r_iv_i\sin\gamma_i=const$. The motion of an object is described by the following equations known from celestial mechanics,
\begin{equation}
\frac{d^2r}{dt^2}-\frac{l^2}{r^3}=-\frac{GM(t,r)}{R^2},
\label{d2rdt2}
\end{equation}
\begin{equation}
\frac{d\phi}{dt}=\frac{l}{r^2},
\label{dfdt}
\end{equation}
where the mass $M(t,r)$ depends on the position of the object in relation to the cloud. In those parts of the trajectory where $r>R(t)$, one has $M(t,r)=M=const$, and for $r\leq R(t)$ one gets $M(t,r)=4\pi\rho(t)r^3/3$.

\begin{figure}
	\begin{center}
\includegraphics[angle=0,width=0.45\textwidth]{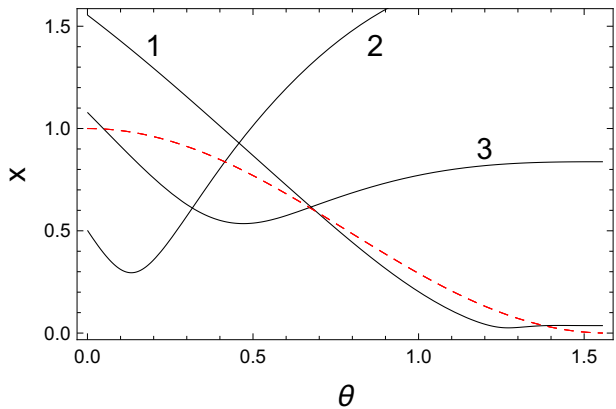}
	\end{center}
\caption{Examples of radial trajectories of planets in the gravitational field of a contracting protostellar cloud. The dashed curve shows the radius of the cloud according to (\ref{rparam}). Curve 1 corresponds to $v_i=0.3$~km~s$^{-1}$, $x_i=1.6$, $\cos\gamma_i=-0.998$. This planet was captured initially. It starts from a large radius, but passes near the center of the cloud at a late stage of collapse and therefore, losing a lot of energy, goes into orbit with a small $a$. The final parameters of its orbit are $a=0.15\times10^3$~au, $e=0.8$. Planet 2 starts from a smaller radius, but is not captured. It corresponds to $v_i=0.75$~km~s$^{-1}$, $x_i=0.5$, $\cos\gamma_i=-0.8$, and final parameters are $a=4.5\times10^3$~au, $e=1.3$. Planet 3 is not captured initially, but is captured in the process of floating through the cloud.  It corresponds to $v_i=0.5$~km~s$^{-1}$, $x_i=1.1$, $\cos\gamma_i=-0.8$, and has the final parameters $a=16\times10^3$~au, $e=0.76$.}
\label{ex}
\end{figure}

For a numerical solution, it is convenient to rewrite the system of equations (\ref{d2rdt2}), (\ref{dfdt}) through the independent variable $\theta$ in the following form:
\begin{eqnarray}
\frac{d^2x}{d\theta^2}+2\frac{dx}{d\theta}{\rm tg}\theta-\frac{\sigma\cos^4\theta}{x^3}=
\nonumber
\\
\nonumber
\\
=\left\{
\begin{array}{ll}
-\frac{2\cos^4\theta}{x^2}; & x\geq \cos^2\theta,
\\
\\
-\frac{2x}{\cos^2\theta}; & x< \cos^2\theta,
\end{array}
\right.
\label{eqbig}
\end{eqnarray}
where denoted $x=r/R_i$ and $\sigma=6t_d^2l^2/R_i^4$.
The equation (\ref{eqbig}) is solved numerically for each statistical test in which the initial values of $x_i$, $v_i$, $\gamma_i$ are played. In the numerical solution, the evolution of the cloud does not continue until $\theta=\pi/2$, but until a slightly earlier moment, at which the radius of the cloud is much smaller than the radius of the Neptune orbit. At this moment the position of the object is $r_f$ and its radial velocity is $dx/d\theta$, recalculated into $v_r=dr/dt$. With a known value of $l=const$, based on these data, the semimajor axis and the eccentricity of the final elliptical orbit around the formed star are calculated:
\begin{equation}
a=\frac{GM}{2|\varepsilon|}, \quad e=\left(1+\frac{2\varepsilon l^2}{(GM)^2}\right)^{1/2},
\end{equation}
where the specific energy is $\varepsilon=(v_r^2+v_t^2)/2-GM/r_f$, $v_t=l/r_f$. Fig.~\ref{ex} shows three characteristic trajectories of the planets. 


\section{Statistics of capturing planets and small objects}

\begin{figure}
	\begin{center}
\includegraphics[angle=0,width=0.45\textwidth]{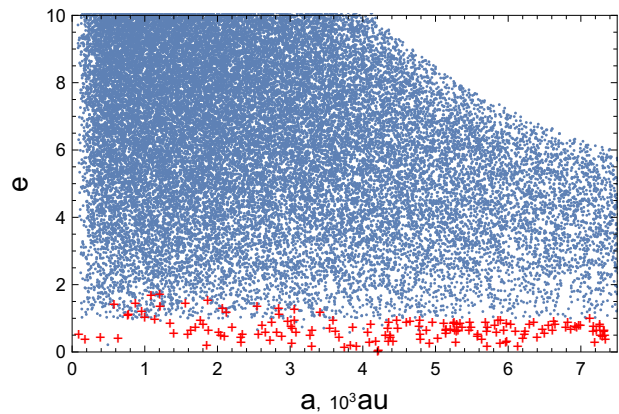}
	\end{center}
\caption{The distribution of events in one statistical simulation on the diagram ``the major semi-axis of the final orbit -- the final eccentricity'', $(a-e)$. The blue circles show planets that were not captured initially, the red crosses show planets whose initial velocity is less than the escape velocity at the same distance from a star with mass $M_\odot$. The major semi-axis is given in units $10^3$~au.}
	\label{gr1}
\end{figure}

\begin{figure}
	\begin{center}
\includegraphics[angle=0,width=0.45\textwidth]{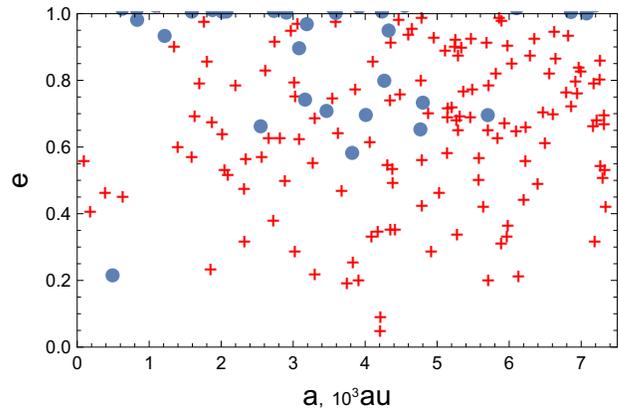}
	\end{center}
\caption{The same as at the Fig.~\ref{gr1}, but for finite orbits with $e<1$.}
	\label{gr2}
\end{figure}

In this paper, seven simulations were performed with the number of statistical tests $4\times10^5$ in each simulation. The statistical variables for which random values are played (taking into account their distribution functions \cite{Stat}) are the initial distance from the center of the protostellar cloud, the initial velocity and the direction of the velocity. Initially, the captured planets belong to the region of star formation, and the protostellar cloud in question is surrounded by other similar clouds at the beginning of the compression, so the effect of gravitational focusing does not play a role in this case. We set the initial velocities and number density of the planets as they should be in a homogeneous star formation region. The characteristic velocity dispersion in such a region is $\sigma_v\simeq1$~km~s$^{-1}$. Distribution of planet velocity 
\begin{equation}
P(v_i)d^3v_i=\frac{1}{(2\pi)\sigma_v^3}\exp\left\{-\frac{v_i^2}{2\sigma_v^2}\right\}d^3v_i,
\label{maksw}
\end{equation}  
where $d^3v_i=2\pi v_i^2dv_id\cos\gamma_i$.

All the results of statistical tests are divided into two groups. In the first group, the initial velocity $v_i<(2GM/r_i)^{1/2}$, i.e. less than the escape velocity for the case of a point mass (these events are shown in Fig.~\ref{gr1} and Fig.~\ref{gr2} with crosses). These objects can be conditionally called initially captured, although not all such objects end up in finite orbits. At the initial moment, they could be inside the cloud at $r_i<R_i$ and, moving away from the center, experience the gravitational influence of only part of the mass of $M$. The second group, shown in Fig.~\ref{gr1} and Fig.~\ref{gr2} circles, corresponds to $v_i\geq(2GM/r_i)^{1/2}$. Despite this initial ratio, some of these objects could be captured in a contracting cloud if they passed close enough to its center and lost a sufficiently large kinetic energy. As expected, the objects of the first group are concentrated mainly near small $e\leq2$.

Consider objects whose final orbit is finite ($e<1$), and the final semimajor axis does not exceed the initial radius of the cloud before the stage of rapid compression, $a<R_i$. Calculations of the number of positive cases in the performed statistical simulations show that such events account for a fraction of $1.6\times10^{-3}$ of all events. Among them, events with $v_i<(2GM/r_i)^{1/2}$ are 87\%, and events with $v_i\geq(2GM/r_i)^{1/2}$ are 13\%. Of particular interest are events with the final parameters of the orbit, approximately corresponding to the assumed parameters of the 9th planet orbit, i.e. with $a<10^3$~au. Such events make up a fraction $3.6\times10^{-4}$ of all events. Among them, events with $v_i<(2GM/r_i)^{1/2}$ are 59\%, and events with $v_i\geq(2GM/r_i)^{1/2}$ are 41\%. Thus, the cloud compression plays a fairly active role, among all the objects captured in the central part (orbits with $a<10^3$~au), almost half of the objects were captured due to cloud compression.

Based on the above simulation results, it is possible to estimate the probability of capturing an additional planet into a contracting cloud. To do this, one needs to know the number of free-floating planets in the star formation region. We use the estimate given in \cite{NapAdaBat21}, where it is assumed that each star loses several Earth masses of solid matter, i.e. approximately one lost planet that becomes free-floating. We also assume that the cloud in question began to shrink rapidly immediately after the fragmentation event (separation from the surrounding similar regions). This corresponds to the fact that the initial volume of $4\pi R_i^3/3$ contains one planet on average. Hence we get that the probability of capture to all finite orbits with $a<R_i$ is 4.4\%. This value is greater than the probability of capturing planets from the Galaxy disk obtained in \cite{GouRib17} ($\sim1$~\%), and is comparable to the probability of capturing a planet in the star formation region into an already formed planetary system ($\sim1-6$~\%) \cite{ParLicQua17}. The probability of capture into sufficiently compact orbits with $a<10^3$~AU according to the results of our simulation is 0.17\%.

The process of capturing into a contracting cloud does not depend on the mass of the captured object (if it is not too large), therefore, it can be expected that not only large planets are captured, but also smaller bodies (asteroids, meteoroids, planetesimals, cosmic dust)  \cite{NapAdaBat21}. This trapped substance is contained mainly in a large volume with a size of the order of the size of the Oort cloud in the solar system. Thus, capture into a protostellar cloud is an additional process of the appearance of cosmic bodies in wide orbits around stars.

In conclusion of this section, we evaluate the role of the gas-kinetic and dynamic friction in the process of capturing an object into the cloud. The characteristic drag time for a planet with a mass $M_p$, a radius of $R_p$ and an initial velocity of $v_i$ is
\begin{eqnarray}
t_{\rm gas}&\sim & \frac{2M_p}{\pi\rho_i v_iR_p}\simeq10^{16}\left(\frac{M_p}{3\times10^{-6}M_\odot}\right)
\nonumber
\\
&\times&\left(\frac{v_i}{1\mbox{~km~s$^{-1}$}}\right)^{-1}\left(\frac{R_p}{6400\mbox{~km}}\right)^{-1}\mbox{~yr},
\end{eqnarray}
where the characteristic parameters of the cloud are taken equal to those given above. This time is comparable to the passage time through the cloud $\sim R_i/v_i$ only for solid particles with sizes less than $R_p\sim10^{-3}$~cm, i.e., the fine cosmic dust is held by the protostellar cloud. Characteristic slowing time of the planet by dynamic friction
\begin{eqnarray}
t_{\rm df}&\sim &\frac{v_i^3}{4\pi G^2\Lambda B\rho_i M_pm}\simeq7\times10^{10}\left(\frac{M_p}{3\times10^{-6}M_\odot}\right)^{-1}
\nonumber
\\
&\times&\left(\frac{v_i}{1\mbox{~km~s$^{-1}$}}\right)^3\mbox{~yr},
\end{eqnarray}
where the Coulomb logarithm is $\Lambda\approx10$, $B\approx0.426$. Consequently, slowing by kinetic and dynamic friction is very weak and cannot contribute to the capture of a planet into the cloud.


\section{Capture of primordial black holes from the Galactic halo}

The velocity distribution of dark matter particles and PBHs in the galactic halo is most often assumed to be isotropic and having the Maxwellian form (\ref{maksw}) with velocity dispersion $\sigma_v\simeq200$~km~s$^{-1}$. In the cloud-related frame, it is also necessary to take into account the velocity shift associated with rotation in the disk around the center of the galaxy. However, due to the large velocity dispersion, the probability of capture at the stage of a contracting protostellar cloud is extremely small. Numerical modeling allowed us to find only the upper limit, less than one statistical test out of $10^6$ leads to capture.  

Let's make a simple analytical estimate of the capturing probability into the cloud for the PBH with a mass of $M_{\rm PBH}$ of the order of the Earth mass. The PBH moves at speed $v_p$ through the cloud during the characteristic time $\Delta t\sim R_i/v_p$. During the dynamic time $t_d$, the gravitational potential inside the contracting cloud changes by an amount of the order of $GM_\odot/R_i$, therefore, the loss of kinetic energy of the passing PBH can be estimated as $\Delta E\sim(GM_{\rm PBH}M_\odot/R_i)\Delta t/t_d$. This loss will exceed the initial kinetic energy of $M_{\rm PBH}v_p^2/2$ in the case 
\begin{equation}
v_p\leq v_{\rm cap}=\left(\frac{2GM_\odot}{t_d}\right)^{1/3}\simeq0.5\mbox{~km~s$^{-1}$}.
\label{vpeq}
\end{equation}
The $v_p$ is of the order of the escape velocity at the cloud boundary. At $v_p\leq v_{\rm cap}$, the PBH will be captured into the cloud. The total number density of PBHs in the Galactic halo in the vicinity of the Sun is $n_{\rm PBH}\sim f\rho_{\rm DM}/M_{\rm PBH}$, where $\rho_{\rm DM}\simeq0.3$~GeV~cm$^{-3}$ is the dark matter density in the region of the Solar system, $f$ is the fraction of the PBHs in the composition of dark matter. If we assume a Maxwellian distribution for the PBHs in the halo (\ref{maksw}) with variance of $\sigma_v\simeq200$~km~s$^{-1}$, then only a small fraction of all the PBHs floating through the cloud with low speeds can be captured. 
During the  compression time $t_d$ of the cloud, it can capture the amount of PBHs
\begin{eqnarray}
\Delta N&\sim& 4\pi R_i^2 n_{\rm PBH}t_d\int\limits_0^{v_{\rm cap}} vP(v)d^3v
\nonumber
\\
&\sim& \frac{(2\pi)^{1/2}R_i^2 n_{\rm PBH}t_dv_{\rm cap}^4}{\sigma_v^3}\simeq4\times 10^{-9}f.
\label{npbh}
\end{eqnarray}
Using the capture cross-section obtained in \cite{NapAdaBat21} (Eq.~(19)) into the already formed solar system, we get that for $5\times10^9$~years, an order of magnitude more PBH will be captured, $\sim4\times 10^{-8}f$, and in \cite{LinLoe18} an even larger value of $\sim10^{-4}-10^{-3}$ was obtained for the probability of the capture. Based on these estimates, we come to the conclusion that the PBH of planetary mass, if it exists in the solar system, was captured not at the stage of the cloud compression, but later for several billion years due to the variable gravitational field of the Sun-Jupiter system of the already formed solar system. At the same time, the number density of low-mass PBHs can be much higher, and the capture of PBHs with a mass $\sim10^{-8}M_\oplus$ into the solar system is very likely.


\section{Conclusion}

The presence of exoplanets with wide orbits, the possible existence of 9th planets in our solar system \cite{BatBro16, BatBro19} and the discovery of a population of free-floating planets \cite{Mroetal17} led to the hypothesis of the capture of free-floating planets in planetary systems due to the variable gravitational field of a star and a giant planet \cite{VarSgaTsi12,GouRib17,ParLicQua17,LiAdam16}. In the case of the solar system, it is the Sun and Jupiter. In \cite{NapAdaBat21}, detailed calculations of the capture cross-section were performed.  

In this paper, we have shown that capture into an already formed planetary system is not the only capture option, but there is also the possibility of capture into a collapsing protostellar cloud at the stage before the formation of a star. Despite the fact that the compression of the cloud in the free fall mode lasts for a short time, during this period there is a strong change in the gravitational potential. In addition, the size of the collapsing cloud is much larger than the size of Jupiter's orbit. Both of these factors lead to the fact that the probability of capture in a collapsing cloud is comparable to the probability of capture in an already formed planetary system.

The capture of free-floating planets is most likely in a protostellar cloud that exists as part of the star formation region, where the number density of lost planets is quite large. According to modern concepts, confirmed by isotope analysis, the solar system at an early stage of its life was in a stellar association formed during the fragmentation of a single giant cloud. Subsequently, this association dissipated as a result of dynamic processes. 

The 9th planet, the evidence of which on the periphery of the solar system was obtained by Batygin et al. \cite{BatBro16, BatBro19}, could be captured both in a planetary system and in a contracting protosolar cloud. It is also possible to capture the PBHs from the Galactic halo into a planetary system or into a contracting cloud. However, the  PBH capture probability is extremely small due to the large velocity dispersion of the  objects in Galactic halo.

This work is supported by the Russian Science Foundation grant 23-22-00013,

https://rscf.ru/en/project/23-22-00013/


\end{document}